\documentclass[prl,aps,reprint,superscriptaddress]{revtex4-2}

\pdfoutput=1

\usepackage[colorlinks=true
,urlcolor=blue
,anchorcolor=blue
,citecolor=blue
,filecolor=blue
,linkcolor=blue
,menucolor=blue
,linktocpage=true
,pdfproducer=medialab
,pdfa=true
]{hyperref}
\usepackage[T1]{fontenc}
\usepackage{fontawesome}
\usepackage{feynmf}
\usepackage{graphicx}
\usepackage{enumitem}
\usepackage{latexsym}
\usepackage{amsfonts}
\usepackage{amssymb}
\usepackage{mathrsfs}
\usepackage{color}
\usepackage{amsmath}
\usepackage{slashed}
\usepackage{dcolumn}
\usepackage{verbatim}
\usepackage{comment}
\usepackage{float}
\usepackage{multirow}
\usepackage{xspace}
\usepackage[normalem]{ulem}
\usepackage{lipsum}

\def\triumf{TRIUMF, 4004 Wesbrook Mall, Vancouver, BC V6T 2A3, Canada}
\def\sfu{Department of Physics, Simon Fraser University, Burnaby, BC V5A 1S6, Canada}
\def\ucr{Department of Physics and Astronomy, University of California, Riverside, CA 92521, USA}

\begin{document}

\title{Probing Lepton Number Violation at Same-Sign Lepton Colliders}
\affiliation{\triumf}
\affiliation{\sfu}
\affiliation{\ucr}

\author{Carlos Henrique de Lima}
\email{cdelima@triumf.ca}
\affiliation{\triumf}

\author{David McKeen}
\email{mckeen@triumf.ca}
\affiliation{\triumf}

\author{John N. Ng}
\email{misery@triumf.ca}
\affiliation{\triumf}

\author{Michael Shamma}
\email{msham008@ucr.edu}
\affiliation{\triumf}
\affiliation{\ucr}

\author{Douglas Tuckler}
\email{dtuckler@triumf.ca}
\affiliation{\triumf}
\affiliation{\sfu}

\begin{abstract}
   Same-sign lepton colliders offer a promising environment to probe lepton number violation. We study processes that change lepton number by two units in the context of Majorana heavy neutral leptons and neutrinophilic scalars at $\mu$TRISTAN, a proposed same-sign muon collider. Our work shows that such colliders, with modest energy and luminosity requirements, can either reveal direct evidence of lepton number violation or significantly constrain unexplored regions of parameter space, especially in the case of a neutrinophilic scalar.
\end{abstract}
\maketitle

\section{Introduction}\label{sec:intro}
More than a decade after the discovery of the Higgs boson, it is still unknown what lies beyond the Standard Model (SM). The Large Hadron Collider (LHC) has provided an unprecedented look at the energy frontier, while the HL-LHC promises to increase the available data by an order of magnitude~\cite{ZurbanoFernandez:2020cco}. To better probe beyond-the-SM (BSM) physics, a broad suite of future colliders have been considered to target the energy frontier or to study the electroweak sector with extreme precision~\cite{P5:2023wyd}. Most studies focus on same-sign hadron or opposite-sign lepton collisions, while same-sign lepton collisions have received less attention. This work highlights the advantages of same-sign lepton colliders in the search for lepton-number--violating (LNV) processes.

In the case of lepton colliders, beams of opposite charge are typically considered, as in the International Linear Collider (ILC)~\cite{Behnke:2013xla,Adolphsen:2013jya,Adolphsen:2013kya,Bambade:2019fyw,Zarnecki:2020ics}, the Compact Linear Collider (CLIC) \cite{Aicheler:2018arh,Zarnecki:2020ics}, the Cool Copper Collider (C$^3$)~\cite{Vernieri:2022fae,Breidenbach:2023nxd}, the HALHF project~\cite{Lindstrom:2023owp,Foster:2023bmq}, and a Muon Collider \cite{Heusch:1995yw,Antonelli:2015nla, Long:2020wfp, Delahaye:2019omf, Accettura:2023ked,MuonCollider:2022nsa}. However, when building these new colliders, it may be advantageous to collide lepton beams of the same sign to understand better how to control and develop the required technology and, along the way, search for new physics. Same-sign lepton collisions have been considered in preliminary studies of $e^-e^-$ collisions at CLIC~\cite{CLICPhysicsWorkingGroup:2004qvu,Schulte:2002ui} and of $\mu^{+}\mu^{+}$ collisions at $\mu$TRISTAN~\cite{Hamada:2022mua}. Collisions with an initial nonzero lepton number have minimal SM background for LNV processes, allowing the probe of unconstrained parameter space of BSM physics with relatively small energy and luminosity compared to the opposite-sign counterpart~\cite{Ananthanarayan:1995cn,Belanger:1995nh,Gluza:1995ix,Fridell:2023gjx,Fukuda:2023yui,Das:2024gfg}. 

LNV is present in most extensions of the neutrino sector~\cite{Proceedings:2019qno,deLima:2022dht,Dvali:2016uhn}, and can provide a compelling explanation for the matter asymmetry of the universe~\cite{Fukugita:1986hr,Fukugita:2002hu,Boyarsky:2009ix,Davidson:2008bu}. A worldwide effort to probe LNV in neutrinoless double beta ($0\nu\beta\beta$) decays is well underway~\cite{Tornow:2014vta,EXO-200:2019rkq,GERDA:2020xhi,CUORE:2021mvw,KamLAND-Zen:2022tow} and can shed light on LNV new physics. At the same time,  $0\nu\beta\beta$ only probes electron-coupled LNV, and new physics could couple stronger to the other flavors. Same-sign lepton colliders are thus an ideal venue to look for LNV in a way complementary to $0\nu\beta\beta$, especially for new lepton-number-charged particles that are too heavy to be produced in nuclear transitions
\begin{figure*}[t!]
     \centering
        \includegraphics[width=0.9\textwidth]{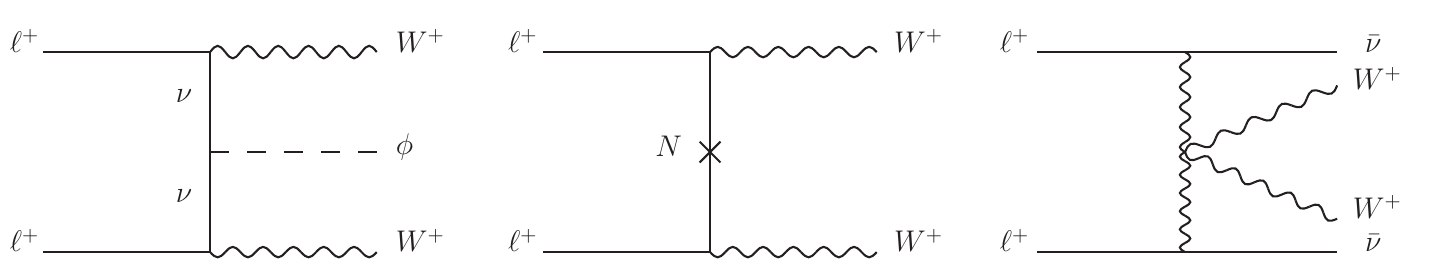}
        \caption{Feynman diagrams for LNV processes via neutrinophilic scalar production (left) and indirect HNL (middle). A representative diagram for the dominant SM background is depicted on the right.}
        \label{fig:diagrams}
\end{figure*}

While LNV at higher scales can be probed at the LHC~\cite{Fuks:2020att,ATLAS:2023tkz,ATLAS:2024rzi,CMS:2024xdq}, proton collisions suffer from large backgrounds that make such searches challenging. It is also possible to probe LNV in an opposite-sign lepton collider~\cite{Mekala:2022cmm,Kwok:2023dck,Li:2023tbx,Mikulenko:2023ezx,Li:2023lkl,Mekala:2023kzo}, but the discovery reach there is restricted by the energy, as the searches can only probe direct production of new LNV states.

The potential to probe LNV in the context of a heavy neutral lepton (HNL) at same-sign lepton colliders was initiated in~\cite{Jiang:2023mte,Santiago:2024zpc,Das:2024kyk}. In this work, we extend the analysis in two important ways. First, we also consider models involving majoron-like scalars carrying two units of lepton number. Second, we leverage the nonzero lepton number of the initial state by considering final states with apparent LNV, which forces the SM background to be significantly smaller as they would contain additional neutrinos carrying away the lepton number. We show that a same-sign lepton collider can probe unexplored regions of parameter space for both LNV models. Our main results are presented in Fig.~\ref{fig:bounds}.

\section{Scenarios with lepton number violation}\label{sec:model}

\subsection{Neutrinophilic Scalars}\label{subsec:nuscalar}

Although an accidental symmetry of the SM, there are many reasons to believe that lepton number is not an exact symmetry at the fundamental level. It may be violated spontaneously when a scalar field with nonzero lepton number acquires a vacuum expectation value (VEV). In this case, at low energies, a physical scalar excitation, $\phi$, with lepton number $L=-2$ couples to the light neutrinos through the interaction
\begin{equation}
{\cal L}\supset\frac12\lambda_{\alpha\beta}\phi\bar \nu_\alpha^c\nu_\beta+{\rm h.c.}
\label{eq:scalareft}
\end{equation}
where $\alpha$ and $\beta$ label the lepton flavor and $\nu^c$ is a charge conjugated neutrino field. The existence of a new scalar like $\phi$ is highly motivated by neutrino mass models with spontaneously broken lepton number~\cite{Gelmini:1980re,Chikashige:1980qk,Barger:1981vd} where it is usually dubbed a Majoron. In addition to their utility in explaining the origin of neutrino masses, Majorons, and similar lepton-number-carrying scalars are well-studied mediators of dark matter (DM) and neutrino self interactions~\cite{DeGouvea:2019wpf,Kelly:2019wow,Kelly:2020aks,Kelly:2021mcd,Benso:2021hhh,Chichiri:2021wvw} and may themselves contribute to the observed DM density~\cite{Heeck:2017wgr,McKeen:2018xyz,Heeck:2019guh}.

The interaction in Eq.~\eqref{eq:scalareft} results from some higher-dimensional operator above the electroweak symmetry breaking scale, e.g. $g_{\alpha\beta}\phi(L_\alpha H)(L_\beta H)/\Lambda^2$ with $L_\alpha=(\nu_\alpha,\ell_{L\alpha})^T$ the SM lepton doublet of flavor $\alpha$, $H$ the Higgs doublet, $\lambda_{\alpha\beta}=g_{\alpha\beta}v^2/\Lambda^2$, and $v=246~\rm GeV$ the Higgs vacuum expectation value (VEV).

In a same-sign lepton collider, the dominant production mechanism of $\phi$ is in association with a pair of same-sign $W$ bosons, as shown by the left diagram of Fig.~\ref{fig:diagrams}. In the high-energy limit, $\sqrt s\gg m_\phi$,~$m_W$, the cross section for this process is
\begin{equation}
\sigma\simeq13~{\rm pb}~\lambda_{\ell\ell}^2\left(\frac{\sqrt s}{1~\rm TeV}\right)^2.
\end{equation}
This cross section is tamed at high energy through, e.g., the dimension-6 operator above~\cite{Worah:1992is}. Relatedly, as we dial $\Lambda$ below the scales involved in the collider production of $\phi$, namely, $m_\phi$ and $\sqrt s$, the operator of Eq.~\eqref{eq:scalareft} should be augmented by other interactions involving $\phi$. Nevertheless, using only this effective operator gives conservative bounds, which is the strategy we use.

At very high energies, production of $\phi$ through vector boson fusion (VBF) becomes important (this process can be seen by rearranging the initial and final states in Fig.~\ref{fig:diagrams} and attaching external lepton legs to the $W$'s). Interestingly, the final state leptons need not be of the same flavor as those in the initial state beams, allowing for probes of different flavor couplings.

Once produced, $\phi$ decays invisibly to neutrinos for $m_\phi<v$. For $m_\phi$ values above the electroweak scale, new decay modes for $\phi$ can open up, with the detailed branching ratios depending on the UV completion~\cite{deGouvea:2019qaz}. We conservatively assume that the primary decay mode for $\phi$ is invisible, although stronger constraints could apply to searches with visible particles in the final state. In what follows, we take a phenomenological approach and do not assume any particular form for the coupling matrix $\lambda_{\alpha\beta}$ motivated by specific neutrino mass models. The production cross sections as a function of $m_\phi$ for $\sqrt{s}=250$ GeV and 1 TeV with $\lambda_{\ell\ell}=1$ are shown in Fig.~\ref{fig:nevents}, respectively. A detailed estimation of the experimental reach is done in the next section.

\subsection{Heavy Neutral Leptons}\label{subsec:HNLs}

Heavy neutral leptons (HNLs), or sterile neutrinos, are common ingredients in neutrino mass models~\cite{Mohapatra:1979ia,Weinberg:1979sa,Schechter:1980gr,Foot:1988aq,Ma:1998dn,Asaka:2005pn,Bondarenko:2018ptm,Abdullahi:2022jlv} and solutions of the matter-antimatter asymmetry of the universe~\cite{Fukugita:1986hr,Fukugita:2002hu,Boyarsky:2009ix,Davidson:2008bu}. An HNL is an SM gauge singlet Weyl fermion, $N$, with lepton number one that interacts with the SM through the neutrino portal operator
\begin{equation}
\mathcal{L} \supset Y_{\alpha} \bar L_\alpha (i\sigma^2 H^\ast) N +{\rm{h.c.}} \, ,
\end{equation}
where $Y_{\alpha}$ is the Yukawa coupling to flavor $\alpha$. After electroweak symmetry breaking, this operator mixes the HNL with SM neutrinos. This induces charged current interactions of the HNL with a charged lepton $\ell_\alpha$ that are suppressed compared to active neutrinos by a factor  $U_{\alpha}$, which is the $\nu_\alpha$-$N$ mixing angle.  
\begin{figure}[t!]
     \centering
        \includegraphics[width=0.495\textwidth]{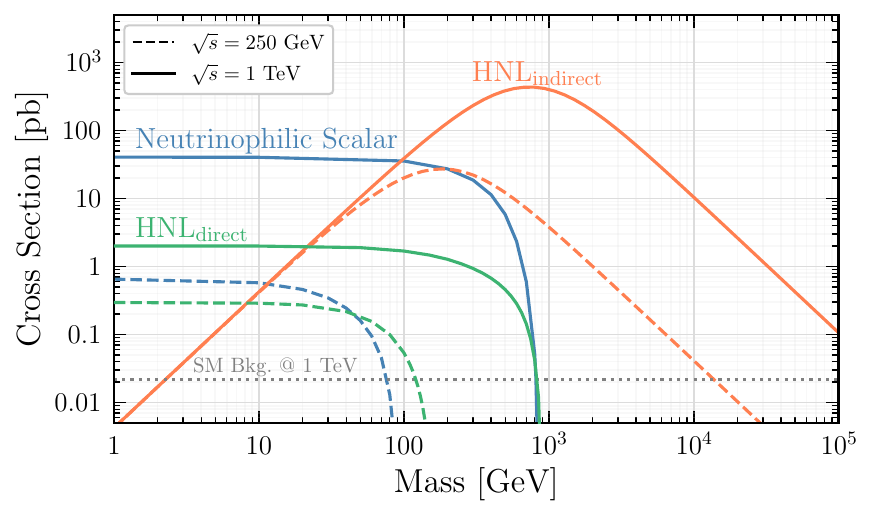}
        \caption{Cross section as a function of mass for the direct production of a neutrinophilic scalar, $\ell^+ \ell^+ \rightarrow W^+ W^+ \phi$ (blue curves), indirect production of a HNL, $\ell^+ \ell^+ \rightarrow W^+ W^+$ (orange curves), and direct HNL production (green curves). The solid (dashed) curves are for $\sqrt{s} = 1$ TeV ($\sqrt{s} = 250$ GeV). We have set $\lambda_{\ell\ell}=1$ in the neutrinophilic scalar model and $\left|U_\ell\right|=1$ in the HNL model. The dotted gray line is the SM background at $\sqrt{s} = 1$ TeV, while for $\sqrt{s} = 250$ GeV, the background has a cross-section of $\sim 10^{-5}$ pb.}
        \label{fig:nevents}
\end{figure}
Since it is a gauge singlet, the simplest way to give $N$ a mass is to allow it to have a Majorana mass that violates lepton number by two units,
\begin{equation}
{\cal L}\supset -\frac12 m_N \bar N^cN+{\rm h.c.}
\end{equation}
In the presence of this mass, processes that violate lepton number by two units can occur via an intermediate HNL, in particular, $\ell^+\ell^+\to W^+W^+$ with a cross-section
\begin{equation}
\sigma\simeq 
\begin{cases}
42~{\rm pb}\left|U_{\ell}\right|^4\left(\frac{m_N}{100~\rm GeV}\right)^2, &m_N\ll\sqrt s
\\
11~{\rm pb}\left|U_{\ell}\right|^4\left(\frac{\sqrt s}{1~\rm TeV}\right)^4\left(\frac{10~\rm TeV}{m_N}\right)^2, &m_N\gg\sqrt s~.
\end{cases}
\end{equation}
In Fig.~\ref{fig:nevents}, we show this cross section as a function of $M$ for $\left|U_\ell\right|=1$ and $\sqrt s=250~\rm GeV$, $1~\rm TeV$. Although in many neutrino mass models the value of $U_\ell$ can be connected to the light neutrino masses, we take an agnostic approach and treat it as a free parameter. We also consider the simplified case of a single HNL mixing with one SM neutrino flavor. As in the neutrinophilic scalar case, the VBF version of the process is important at high energies and allows couplings to lepton flavors not present in the beam to be probed. 

There is also the possibility of producing the HNL directly at same-sign lepton colliders. The dominant production mode for $\sqrt s\lesssim {\rm TeV}$ is $\ell^+ \ell^+ \rightarrow \ell^+ \, W^{+} \, N$, while at higher energies the VBF process $\ell^+ \ell^+ \rightarrow W^{+} \, W^{+} \, N \, \bar{\nu}_{l}$ can become significant. The HNL then decays to electroweak bosons and leptons. The Majorana nature of HNLs allows them to decay to both positively- and negatively-charged leptons, with the LNV mode $N \to W^+ \ell^-$ occurring 30\% of the time for $m_N\gg m_W$.

Even though a same-sign collider works best for indirect probes of Majorana HNLs, a competitive reach for direct searches is still possible. The following section provides a detailed estimation of the discovery reach of both the direct and indirect processes.

\begin{figure*}[t!]
     \centering
        \includegraphics[width=0.95\textwidth]{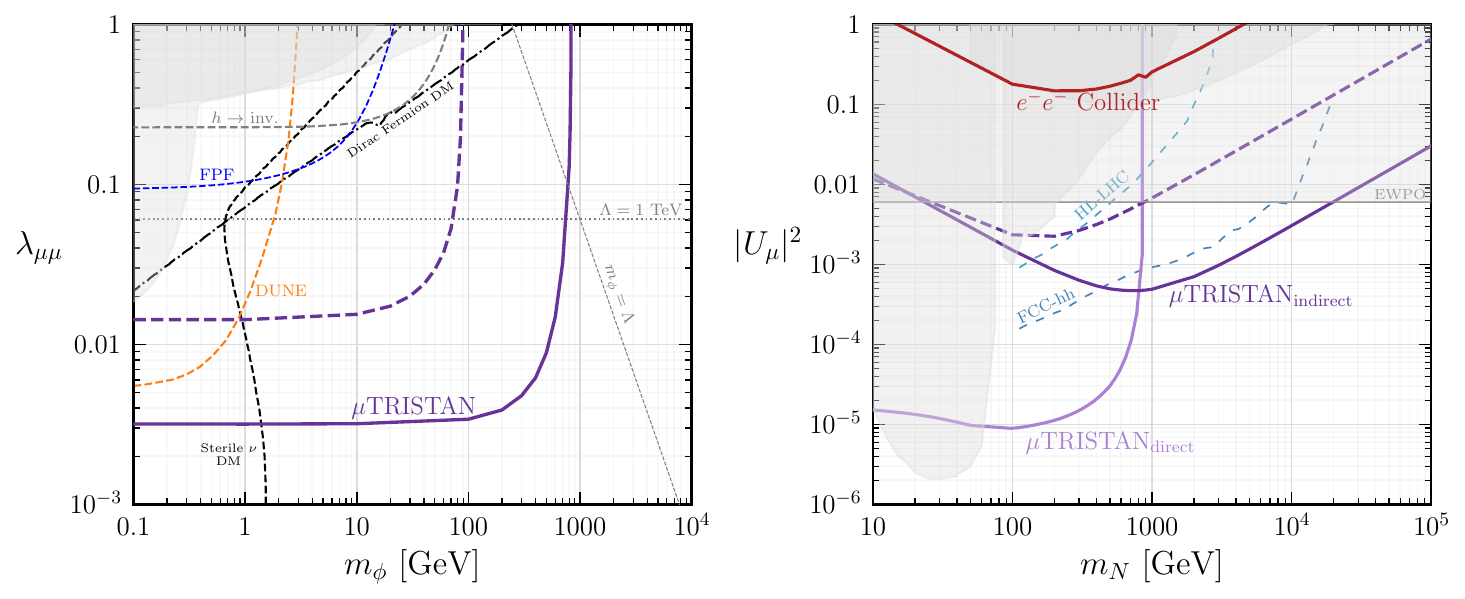}
        \caption{Sensitivity of $\mu$TRISTAN (purple curves) and $e^- e^-$ collider (red curves) to neutrinophilic scalars (left) and HNLs (right) based on a signal rate analysis. The solid (dashed) curves are for $\sqrt{s} = 1$ TeV ($\sqrt{s} = 250$ GeV) with an integrated luminosity of 1 ab$^{-1}$ (100 fb$^{-1}$). The gray-shaded regions are existing constraints; see text for details.}
        \label{fig:bounds}
\end{figure*}

\section{Same sign collider analysis}\label{sec:analysis}
The key signature of both models is the (apparent) LNV process $\ell^+ \ell^+ \to W^+ W^+ (+\phi)$. Lepton number is violated due to the absence of same-flavor-anti-neutrinos $\bar{\nu}_\ell$ in the final state. Since $\phi$ has lepton number $L = -2$, this is only an apparent LNV, while explicitly violated in the HNL mediated process. The advantage of a same-sign collider is the presence of a nonzero initial lepton number. The production of same-sign $W$ bosons in the SM must be accompanied by anti-neutrinos with the same flavor as the incoming beams. 

We focus on muon-neutrino--coupled new physics in the models described above at $\mu$TRISTAN, a recently proposed high-energy $\mu^+ \mu^+$ collider at JPARC~\cite{Hamada:2022mua}. We study hadronic $W$ decays, resulting in a four jet ($4j$) final state. The main SM background is $\mu^{+} \mu^{+} \rightarrow 4j + \bar{\nu}_\ell\bar{\nu}_\ell $ where the dominant contribution is from $\ell^+ \ell^+ \to W^+ W^+ \bar{\nu}_\ell\bar{\nu}_\ell$ as depicted in the right diagram of Fig.~\ref{fig:diagrams}. There is also a subdominant background from $\ell^+ \ell^+ \to Z W^+ \ell^{+}\bar{\nu}_\ell$ where the muon is undetected, mimicking $4j$ plus missing energy.

For both model hypotheses, we consider two energy and luminosity stages: (i) $\sqrt{s} = 250$ GeV with an integrated luminosity of 100 fb$^{-1}$ and (ii) $\sqrt{s} = 1$ TeV with an integrated luminosity of 1 ab$^{-1}$. We generate events using  \texttt{MadGraph5\_aMC@NLO}~\cite{Alwall:2014hca} interfaced with \texttt{Pythia8.2}~\cite{Sjostrand:2014zea} for showering and hadronization, and estimate the detector efficiencies using  \texttt{Delphes}~\cite{deFavereau:2013fsa} with the CLIC card~\cite{Leogrande:2019qbe}. We reconstruct jets in both stages using the Valencia Linear Collider (VLC)~\cite{Cacciari:2011ma,Boronat:2016tgd,Boronat:2014hva} algorithm in exclusive mode for jet clustering with a cone size parameter $R = 1$. The  experimental sensitivity is estimated by using the median expected exclusion significance~\cite{Cowan:2010js,Kumar:2015tna}
\begin{equation}
Z_\text{excl} = \sqrt{2(s-b \ln(1+s/b))}
\end{equation}
where $s$ and $b$ are the number of signal and background events, respectively. We require $Z_\text{excl} > 1.645$ for 95\% C.L. exclusion.

\subsection{Neutrinophilic Scalars}
For the neutrinophilic scalar search, we select events containing four jets with transverse momentum greater than $20~\rm GeV$ and no leptons with transverse momentum larger than $5~\rm GeV$. These simple cuts are sufficient to obtain strong bounds. Further improvements could be made by analyzing the missing energy distribution of the signal and background.

We show the sensitivity of same-sign lepton colliders in the left plot of Fig.~\ref{fig:bounds} for muon-coupled neutrinophilic scalars. Existing constraints from meson decays and $Z$ invisible decays are shown by the gray shaded regions~\cite{Berryman:2018ogk,Brdar:2020nbj,Dev:2024twk}, while future projections from invisible Higgs decays at HL-LHC~\cite{Cepeda:2019klc,deGouvea:2019qaz} are depicted by the dashed gray line. The dashed orange and blue curves are the expected sensitivity of DUNE~\cite{Kelly:2019wow} and the Forward Physics Facility (FPF)~\cite{Kelly:2021mcd}, respectively. Additionally, we show where the scalar mass is the same as the scale of the dimension-6 operator by the gray dashed line labeled $m_\phi = \Lambda$, setting the coupling $g_{\mu\mu}=1$. Above this line our EFT treatment is conservative, and additional signatures could be present that offer stronger bounds. Similarly, we indicate where $\Lambda=1~\rm TeV$, above which a similar statement applies for searches with $\sqrt s=1~\rm TeV$.

With $\sqrt{s} = 250$ GeV, we see that $\mu$TRISTAN probes new regions of parameter space beyond existing bounds and HL-LHC projections. At low masses, $m_\phi \lesssim 1$ GeV, $\mu$TRISTAN surpasses the expected sensitivity of the FPF, while for higher masses, $\mu$TRISTAN outperforms both the FPF and DUNE. Increasing the energy to $\sqrt{s} = 1$ TeV, we see that $\mu$TRISTAN probes neutrinophilic scalars up to $m_\phi \simeq 800$ GeV, far beyond the sensitivity of DUNE and LHC experiments. In addition, $\mu$TRISTAN fully probes the relic abundance lines of self-interacting sterile neutrino dark matter \cite{DeGouvea:2019wpf,Kelly:2020aks,Kelly:2021mcd} and $\phi$-mediated Dirac fermion dark matter \cite{Kelly:2019wow,Kelly:2021mcd} benchmarks, depicted by the black dashed and dot-dashed curves, respectively. Overall, we see that a same-sign muon collider like $\mu$TRISTAN has an unprecedented sensitivity to neutrinophilic scalars.

\subsection{Heavy Neutral Leptons}
The Majorana HNL analysis has the same requirements as the neutrinophilic scalar case, with the addition of a cut on the missing transverse energy, $E_T^\text{miss} < 20~\text{GeV}$, since the signal has no missing energy. This enforces that we are probing actual LNV and significantly improves the reach for this model. This feature can be used to easily distinguish the HNL vs the neutrinophilic scalar in case of a measurement different from the SM. 

In the right plot of Fig.~\ref{fig:bounds}, we show the sensitivity of same-sign lepton colliders to HNLs. The reach of $\mu$TRISTAN at $\sqrt{s} =250$ GeV with a luminosity of 100 fb$^{-1}$ is shown by the dashed purple curve. We see that the reach goes beyond existing ATLAS and CMS searches \cite{ATLAS:2023tkz,CMS:2024xdq} (gray shaded regions), as well as HL-LHC projected sensitivity \cite{Pascoli:2018heg} for $m_N \gtrsim 300$ GeV. Increasing the center-of-mass energy and luminosity to $1~\rm TeV$ and $1~\rm ab^{-1}$, respectively, pushes the reach to higher masses and is competitive to FCC-hh projections \cite{Pascoli:2018heg}. Furthermore, $\mu$TRISTAN goes beyond constraints from electroweak precision observables depicted by the horizontal gray line at $|U_\mu|^2 \sim 0.01$ \cite{Bertoni:2014mva}.

We also show the sensitivity of an $e^- e^-$ collider with the red curve in Fig.~\ref{fig:bounds}. The analysis, in this case, requires final state leptons of a different flavor, with the dominant background being the same as before, but with the $W^{+} \rightarrow \mu \nu_{\mu}$. The reach, in this case, is weaker, but it is a proof of concept to probe different flavor couplings. This opens the possibility of probing, for example, tau-coupled HNLs in the same sign muon collider.

The sensitivity of $\mu$TRISTAN for direct HNL production is depicted by the light-purple curve in Fig.~\ref{fig:bounds}. Because the initial beams are anti-muons, we require the HNL to decay to a muon $N \to W^+ \mu^-$. The signature in this case is $\mu^+ \mu^+ \to W^+ W^+ \mu^+ \mu^-$ and violates lepton number by two units. Again, we consider hadronic $W$ decays and select final states with two opposite sign muons with $p_T^{\ell} > 20~\text{GeV}$, 4 jets that each have $p_{T}^{j} > 20~\text{GeV}$ and no additional leptons. In the SM, the main background has additional missing energy as it comes from  $\mu^+ \mu^+ \to W^+ W^+ \mu^+ \mu^-\bar{\nu}_\mu \bar{\nu}_\mu$. Therefore, we require $E_T^\text{miss}  < 20~\text{GeV}$ since the signal has no missing energy.  We see that $\mu$TRISTAN probe regions of parameter space complementary to existing LHC searches and future sensitivity of HL-LHC.

\section{Discussion}\label{sec:conclusion}

This work has explored the unique advantages of same-sign lepton colliders in probing LNV, a phenomenon that can provide critical insights into BSM physics. We focused on two promising scenarios: the production of neutrinophilic scalars and heavy neutral leptons (HNLs). Both models offer signatures that can be effectively explored in a same-sign lepton collider environment. Notably, the presence of a nonzero initial lepton number significantly reduces the SM background, enhancing the sensitivity of these colliders to LNV processes. This allows for a clearer detection of events that would otherwise be obscured in traditional colliders. We have shown that obtaining experimental sensitivity that is inaccessible to other current and future experiments is possible. This is especially the case for a neutrinophilic scalar with mass $1~{\rm GeV}\lesssim m_\phi\lesssim 800~{\rm GeV}$ and an HNL with $m_N\gtrsim 100~\rm GeV$. 

While neutrinoless double beta decay experiments are crucial for studying LNV, our findings advocate for a complementary approach with same-sign lepton colliders. These colliders can explore a broader range of lepton flavors and provide a unique experimental environment, further enriching our understanding of the neutrino sector and its potential extensions.


\acknowledgments

We thank M. Swiatlowski and the participants at the ``Physics Potential of Future Colliders'' Workshop at TRIUMF for helpful discussions. We thank Deepak Sathyan for providing data tables for updated constraints on neutrinophilic scalars. This work is supported by Discovery Grants from the Natural Sciences and Engineering Research Council of Canada (NSERC). TRIUMF receives federal funding via a contribution agreement with the National Research Council (NRC) of Canada. MS is supported by the US Department of Energy under award number DE-SC0008541.

\bibliographystyle{utphys}
\bibliography{ref}

\end{document}